\documentstyle[preprint,prd,aps,epsf]{revtex}
\newcommand{\NP}{Nucl. Phys. }
\newcommand{\PR}{Phys. Rev. }

\begin{document}
\thispagestyle{empty}
\draft
\tighten
\title{
\hfill {\small MSUHEP-70912} \\
\hfill {\small TUIMP-TH-97/89}\\ 
\hfill \\
Production of neutral scalar Higgs bosons at $\bbox{e\gamma}$ colliders}
\author{Yi Liao$^{1,2}$ and Wayne W. Repko$^3$}
\address{$^1$China Center of Advanced Science and Technology (World Laboratory),
\\ Beijing 100080, P. R. China, \\
$^2$Department of Modern Applied Physics, Tsinghua University,
Beijing 100084, P.R.China \\
$^3$Department of Physics and Astronomy, \\ Michigan State University,
East Lansing, Michigan 48824}
\date{\today}
\maketitle

\begin{abstract}
We study the production of neutral scalar ($CP$ even) Higgs bosons in the
process $e\gamma\to e h$ by including supersymmetric corrections to the 
dominant $t$-channel photon exchange amplitude. In addition to the standard 
model $W^{\pm}$ and fermion
loops, there are substantial contributions from chargino loops. For some cases, 
these contributions can exceed those of the $W$'s and ordinary fermions.
The cross sections in this channel are generally one or two orders of 
magnitude larger than those in the related channel $e\bar{e}\to\gamma h$.
\end{abstract}

\section{Introduction}

In general, extensions of the Standard Model (SM) have a Higgs sector 
consisting of several Higgs bosons in addition to the neutral Higgs associated
with the minimal symmetry breaking doublet. The direct detection of this rich 
spectrum of Higgs bosons at future colliders is a crucial test of 
such models \cite{hunter}. While hadron colliders generally have a larger
luminosity and center of mass energy, lepton colliders offer a much cleaner
environment in which to explore the Higgs sector. In this paper, we study the 
production of the neutral scalar Higgs bosons $h^0,~H^0$ in the minimal 
supersymmetric standard model (MSSM) at $e^+ e^-$ colliders operating in the
$e~\gamma$ mode. This extends the work in Ref. \cite{dicus}, where
the production of the pseudoscalar Higgs boson $A^0$ was investigated.
The related process $h\gamma$ associated production at $e^+~e^-$
colliders was previously studied for the SM Higgs boson in
Refs.\cite{barroso,abbasabadi} and for the MSSM Higgs bosons
in Ref.\cite{djouadi}. The basic observation used here is
that the $t$-channel photon in the $e~\gamma$ mode can approach the physical
region $t=0$, so that the cross section is significantly
enhanced compared to that in the $e^+~ e^-$ mode \cite{dicus}.
This makes the processes $e\gamma\to e~(H^0,~h^0,~A^0)$ 
potentially important channels for the study of Higgs bosons.

In the next Section, we derive the amplitudes and cross sections for the various
$e\gamma\to e\,h$ channels. This is followed by a detailed discussion of the
MSSM and SM Higgs-boson production cross sections using a selection of values
for the model parameters. The MSSM Higgs-boson couplings are contained in the
Appendix.

\section{Amplitudes and cross sections for Scalar Higgs Production}

In the lowest nontrivial order, the process $e^- \gamma\to e^- h$ proceeds
through $t$-channel $\gamma^{\star}(Z^{\star})\gamma h$ triangle diagrams,
box diagrams and associated $s$-channel $e^+ e^- h$ triangle diagrams.
Using the notation of Ref. \cite{hunter}, 
$h$ denotes any of the neutral scalar Higgs bosons, $h^0$ denotes the 
lighter one and $H^0$ the heavier one in MSSM, and the SM Higgs boson is 
denoted by $\phi^0$. As shown in Ref. \cite{dicus}, the dominant 
contribution comes from $t$-channel
$\gamma^{\star}\gamma h$ triangles and the other diagrams can
be safely ignored. We work in the same approximation.
In addition to ordinary fermions and charginos $\chi^{\pm}_i ~(i=1,2)$ which
are the only particles present in the $\gamma^{\star}\gamma A^0$ triangle,
the particles in the $\gamma^{\star}\gamma h$ triangle
can also be $W^{\pm}$ (and associated pseudo-Goldstone bosons
$G^{\pm}$ and Faddeev-Popov ghosts $\eta^{\pm}$), charged Higgs bosons
$H^{\pm}$, and sfermions $\tilde{f}$.
In the conventional approach, the $W^{\pm}$ contribution to the
$\gamma^{\star}\gamma h$ triangle is not gauge invariant, and gauge
invariance is recovered only when all $W^{\pm}$ contributions are
summed. Thus, to make the $t$-channel $\gamma^{\star}$ dominance
self-consistent, it is necessary that a gauge invariant method be
used to compute the $W^{\pm}$ contribution. For this
purpose we can use the background field method \cite{bfm} or
nonlinear gauges \cite{nonlinear}. These methods have the additional
advantage of reducing the number of Feynman diagrams.

The amplitude for the process
$e^-(p)\gamma(\epsilon(k_1),k_1)\to e^-(p')h(q)$ can be written as
\begin{equation}
{\cal A}=-\frac{2\alpha^2 m_W}{\sin\theta_W}\frac{1}{t}
\epsilon^{\mu}(k_1)\bar{u}(p')\gamma^{\nu}u(p)
\left[(k_{2\mu}k_{1\nu}-g_{\mu\nu}k_1\!\cdot\! k_2)T_1(t,m^2_h)
+i\epsilon_{\mu\nu\alpha\beta}k_1^{\alpha}k_2^{\beta}T_2(t,m^2_h)
\right],
\end{equation}
with
\begin{eqnarray}
T_1(t,m^2_h) & = & \lambda(W^{\pm})\left[8C(W^{\pm})+(3+\frac{m^2_h}{2m^2_W})
C'(W^{\pm})\right] -\lambda(f)N_fQ^2_f\frac{2m^2_f}{m^2_W}[C(f)+C'(f)]
\nonumber \\ [4pt]
&   &+\lambda(H^{\pm})C'(H^{\pm}) - \lambda(\chi^{\pm}_i)
\frac{4m_{\chi^{\pm}_i}}{m_W}\left[C(\chi^{\pm}_i)+C'(\chi^{\pm}_i)\right]
-\lambda(\tilde{f})N_{\tilde{f}}Q^2_{\tilde{f}}C'(\tilde{f})\,,\\
[4pt]
T_2(t,m^2_h) & = &-\lambda'(\chi^{\pm}_i)
\frac{4m_{\chi^{\pm}_i}}{m_W}C(\chi^{\pm}_i)\,,
\end{eqnarray}
and a summation over flavors of the fermions $f$, the sfermions
$\tilde{f}$, and the two charginos $\chi^{\pm}_i (i=1,2)$ is implied.
Here, $k_2$ is the momentum of the $t$-channel virtual photon with
$k_2\,=\,p-p'=q-k_1$ and $k_2^2\,=\,t\leq 0$. $N_f$ and $Q_f$
( $N_{\tilde{f}}$ and $Q_{\tilde{f}}$ ) are, respectively, the number
of colors and the electric charge of fermion $f$ ( sfermion $\tilde{f}$ ), and
the $\lambda(f)$'s are reduced couplings of loop particles $f$ to $h$, which are
derived from Ref. \cite{hunter} and given in the Appendix.

By using the background field method for the $W^{\pm}$ loop,
all loop integrals are combinations of two scalar functions
$C(f)$ and $C'(f)$, which depend only the particle type $f$. Specifically, $C(f)
$ and $C'(f)$ are defined by
\begin{eqnarray}
& &\int\frac{d^4 l}{(2\pi)^4}
\frac{1}{(l^2-m^2_f)[(l-k_1)^2-m^2_f][(l+k_2)^2-m^2_f]}
 =\frac{i}{(4\pi)^2}C(f)
\\ [4pt]
& &\int\frac{d^4 l}{(2\pi)^4}
\left[\frac{(2l-k_1)_{\mu}(2l+k_2)_{\nu}}
           {(l^2-m^2_f)[(l-k_1)^2-m^2_f][(l+k_2)^2-m^2_f]}
-\frac{g_{\mu\nu}}{[(l-k_1)^2-m^2_f][(l+k_2)^2-m^2_f]}
\right] \nonumber\\
& &= \frac{i}{(4\pi)^2}
[(k_{2\mu}k_{1\nu}-g_{\mu\nu}k_1\!\cdot\! k_2)C'(f)
+(k_{1\mu}, k_{2,\nu}~~{\rm terms})]
\end{eqnarray}
where in Eq.\,(5), the $k_{1\mu}$ terms (including
$k_{1\mu}k_{1\nu}$ and $k_{1\mu}k_{2\nu}$) do not contribute to
the physical amplitude due to $k_1\!\cdot\!\epsilon(k_1)=0$ and
the $k_{2\mu}k_{2\nu}$ term vanishes using currrent conservation.
For $k^2_1=0$, the function $C(f)$ reduces to an
elementary function \cite{hunter,abbasabadi}, while
$C'(f)$ can be expressed in terms of $C(f)$ and the difference of two-point 
functions as
\begin{eqnarray}     
C(f) & = &\frac{1}{q^2-t}\left[G\left(\frac{q^2}{4m^2_f}\right)
                         -G\left(\frac{t}{4m^2_f}\right)\right], \\ [4pt]
C'(f) & = &-\frac{1}{q^2-t}\left[2+4m^2_f C(f)
+\frac{2t}{q^2-t}\left(F\left(\frac{q^2}{4m^2_f}\right)-
F\left(\frac{t}{4m^2_f}\right)\right)\right]\,,
\end{eqnarray}
where $q^2=(k_1+k_2)^2 = m^2_h,~t = k^2_2$. The scalar functions $F(\tau)$ and
$G(\tau)$ are
\begin{equation}
\displaystyle{
F(\tau)=\left\{ \begin{array}{lc}
2\left[1-\sqrt{1-\displaystyle\frac{1}{\tau}}
 \ln(\sqrt{1-\tau}+\sqrt{-\tau})\right], &~\tau\leq 0 \\
& \\
2\left[1-\sqrt{\displaystyle\frac{1}{\tau}-1}\arcsin\sqrt{\tau}\right], 
&~0\leq\tau\leq 1 \\
& \\
2\left[1-\sqrt{1-\displaystyle\frac{1}{\tau}}
\left(\ln(\sqrt{\tau-1}+\sqrt{\tau})-i\displaystyle\frac{\pi}{2}\right)\right], 
    &~\tau\geq 1
                \end{array}\right.
             }
\end{equation}
and
\begin{equation}
\displaystyle{
G(\tau)=\left\{ \begin{array}{lc}
2\left[\ln(\sqrt{1-\tau}+\sqrt{-\tau})\right]^2, &~\tau\leq 0 \\
& \\
-2\left[\arcsin\sqrt{\tau}\right]^2, &~0\leq\tau\leq 1 \\
& \\
2\left[\ln (\sqrt{\tau-1}+\sqrt{\tau})-i\displaystyle\frac{\pi}{2})\right]^2, 
&~\tau\geq 1.
\end{array}\right.\,,
             }
\end{equation}

The spin-averaged differential cross section is
\begin{equation}
\frac{d \sigma(e\gamma\to eh)}{d(-t)}=
\frac{\alpha^4 m^2_W}{16\pi\sin^2\theta_W}
\frac{s^2+(m^2_h-s-t)^2}{(-t)s}
\left(|T_1(t,m^2_h)|^2+|T_2(t,m^2_h)|^2
\right)\,,
\end{equation}
with $s=(k_1+p)^2=2k_1\!\cdot p$. Due to the logarithmic singularity
$\ln (-t)$ in the forward direction, an angular cutoff is imposed
on the scattered electron $\theta\geq\theta_{\rm min}$,
which is equivalent to the constraint
$t\leq t_{\rm max}=-\eta(s-m^2_h)$, with
$\eta=\sin^2(\theta_{\rm min}/2)$. To obtain the experimentally
measured quantity, we take the convolution of the above cross section
with the photon distribution function $F_{\gamma/e}(x)$,
\begin{eqnarray} \label{sig}
\sigma_{\rm total}(s)& = &\frac{\alpha^4 m^2_W}{16\pi \sin^2\theta_W}
\int^{x_{\rm max}}_{m^2_h/s} dx F_{\gamma/e}(x)
\int^{xs-m^2_h}_{\eta(xs-m^2_h)}\frac{d(-t)}{(-t)}
\left[2-2\frac{m^2_h-t}{xs}+\left(\frac{m^2_h-t}{xs}\right)^2\right]
\nonumber \\ [4pt]
&   &\times\left[|T_1(t,m^2_h)|^2+|T_2(t,m^2_h)|^2\right]\,,
\end{eqnarray}
where, for backscattered laser photons \cite{ginzburg}, $x_{\rm max} = 0.83$ and
\begin{eqnarray}
F_{\gamma/e}(x) & = &\frac{1}{D(\xi)}
\left[1-x+\frac{1}{1-x}-\frac{4x}{\xi (1-x)}
     +\left(\frac{2x}{\xi(1-x)}\right)^2\right]\,\\[4pt]
D(\xi) & = &\left(1-\frac{4}{\xi}-\frac{8}{\xi^2}\right)\ln(1+\xi)
+\frac{1}{2}+\frac{8}{\xi}-\frac{1}{2(1+\xi)^2},~~\xi=2(1+\sqrt{2})\,.
\end{eqnarray}

\section{Discussion}

Before discussing numerical results let us give a brief description
of input parameters. The SM parameters are taken from Ref. \cite{pdg},
$\alpha=1/128,~m_W=80.33$ GeV, $m_Z=91.187$ GeV, $\cos\theta_W=m_W/m_Z$.
In the fermionic contribution, only $t,~b$ and $\tau$ are included using
the masses $m_t=176$ GeV, $m_b=4.5$ GeV, and $m_{\tau}=1.777$ GeV.
In MSSM, the Higgs sector contains four masses
($m_{H^0},~m_{h^0},~m_{A^0},~m_{H^{\pm}}$) and two angles
($\alpha,~\beta$), but only two of these are independent. As usual, we
take $\beta$ and one of the neutral Higgs masses as free parameters.
Special care must be taken when $\beta$ and $m_{h^0}$ are taken free,
since the tree level relations between masses and angles require that
$\tan^2 \beta >(1+\sqrt{r_h})/(1-\sqrt{r_h})$ or
$\tan^2 \beta <(1-\sqrt{r_h})/(1+\sqrt{r_h})$ with
$r_h=m^2_{h^0}/m^2_Z$. In the chargino sector, the mixing pattern is
determined by $\beta$ and two new parameters $M,~\mu$. For the sake
of simplicity we assume that $M$ and $\mu$ are real. Then the
diagonalizing matrices $U$ and $V$ are real and orthogonal,
and CP is conserved in the chargino couplings. The CP-violating form
factor $T_2$ vanishes correspondingly due to
$\lambda'(\chi^{\pm}_i)=0$. The sfermion sector is generally afflicted
with many new parameters. Again for simplicity we assume
$M_{\tilde{Q}}=M_{\tilde{u}}=M_{\tilde{d}}$,
$M_{\tilde{L}}=M_{\tilde{\nu}}=M_{\tilde{e}}$, where
$\tilde{Q},~\tilde{u}$ and $\tilde{d}$ denote the left-handed
squark doublet, the right-handed up-type and down-type squark singlet,
and similarly for $\tilde{L},~\tilde{\nu}$ and $\tilde{e}$.
Without taking into account the mixing between left-handed and
right-handed sfermions, their masses are then expressed in terms of
the above masses $M_{\tilde{f}}$ and $\beta$. Only the mixing
between $\tilde{t}_L$ and $\tilde{t}_R$ is included, which introduces
another parameter $A_t$ from the soft supersymmetry-breaking potential.
To summarize, we have the following free parameters:
$\beta$, one of $m_{H^0,h^0,A^0}$, $M$, $\mu$, $A_t$,
$M_{\tilde{Q},\tilde{u},\tilde{d}}$ and
$M_{\tilde{L},\tilde{\nu},\tilde{e}}$.

The results based on Eq.\,(\ref{sig}) are plotted in Figs.\,1-4 for MSSM Higgs
bosons and in Fig.\,5 for the SM Higgs boson. The angular cutoff is
chosen to be $\eta=10^{-5}$ except for Fig.\,3 in which $\eta=10^{-4}$.

In Fig.\,1, we show the production cross sections for the light Higgs boson 
$h^0$ as a function of its mass $m_{h^0}$ and the center of mass energy 
$\sqrt{s}$. Different choices of $M_{\tilde{f}},~M,~\mu$ and $A_t$ have only a
marginal effect on the rates. This is because the dominant contribution
comes from the standard $W^{\pm}$ and fermion loops. These are much larger
than contributions from charginos and especially sfermions.
The value of $\tan\beta$ affects production rates significantly for
a relatively light $h^0$ with $m_{h^0}<75$ GeV. As $m_{h^0}$ approaches
its tree level upper limit $m_Z$, the rates become insensitive to $\tan\beta$
since in this limit the reduced coupling of $h^0$ to $W^{\pm}$ is
very close to unity for all sufficiently large $\tan\beta$.
The production cross sections are relatively flat as $\sqrt{s}$ varies from 
$200$ GeV to $1000$ GeV, something also seen in the other figures. This 
behaviour is very different from that of associated photon-Higgs production
in the $e^{+}e^-$ mode$\cite{barroso,abbasabadi,djouadi}$,
and can be attributed to a balance between two opposite effects.
On the one hand, the $t-$channel photon carries a momentum
of the scale $\sqrt{s}$ thus tending to suppress rates at high energies. On the
other, the scattered electrons are more concentrated around the forward cone
at higher energies so that the rates are enhanced by an almost on-shell
$t-$channel photon. This may be verified by choosing a larger value of
$\eta$ which is equivalent to subtracting out a larger cone around the
forward direction. For example, using the same input parameters as the
solid curve in Fig.\,1(c) except that $\eta=10^{-2}$, the cross section 
decreases by about a factor of eight between $\sqrt{s}=200$ GeV and 
$\sqrt{s}=1000$ GeV.

The production cross sections for the heavy Higgs boson $H^0$ are shown in
Fig.\,2 as a function of its mass $m_{H^0}$ and $\sqrt{s}$. In this case,
different choices of $M_{\tilde{f}},~M,~\mu$ and $A_t$ have a significant 
impact on rates especially for $m_{H^0}>250$ GeV. This can be understood
by noting that, compared with the case of $h^0$, the dominant contribution 
from $W^{\pm}$ and fermions is relatively suppressed and the other 
contributions, which are sensitive to these parameters, become equally
important. As $m_{H^0}$ moves down to $100$ GeV, the four curves tend
to approach one another. Besides the flatness in $\sqrt{s}$  discussed 
above, this is also due to the fact that the reduced coupling 
$\lambda(W^{\pm})$ happens to be almost the same for $\tan\beta=2$ and $20$ 
when $m_{H^0}=100$ GeV.

Contributions from different particles are displayed in Figs.\,3 (a) and (c)
for the production of $h^0$, and in (b) and (d) for $H^0$. We find that
contributions from $H^{\pm}$ and $\tilde{f}$ loops are always very small 
and can thus be safely ignored. Consider first Figs.\,3 (a) and (c).
As mentioned above, the dominant contribution to $h^0$ production arises
from $W^{\pm}$ and fermion loops. While the $\chi^{\pm}_i$ contribution
is clearly distingushable, it is several times smaller than those from the
$W$'s and fermions. The two contributions interfere destructively
for $m_{h^0}<85$ GeV, reminiscent of the cancellation between $W^{\pm}$
and ordinary fermion loops. For our choice of input parameters,
the reduced coupling $\lambda(\chi^{\pm}_i)$ vanishes around $m_{h^0}=85$
GeV. Then, as $m_{h^0}$ continues to increase, the sign of 
$\lambda(\chi^{\pm}_i)$ is reversed so that the two contributions interfere 
constructively. In Fig.\,3 (b) and (d) the SM and chargino contributions are 
of the same order of magnitude. For large enough $m_{H^0}$, $\chi^{\pm}_i$
can even dominate over $W^{\pm}$ and fermions (though at that point the total
cross section is lower by one or two orders of magnitude), since the once 
dominant role played by $W^{\pm}$ is diminished by its very small coupling 
$\lambda(W^{\pm})$. We also see that the two contributions always 
interfere constructively. This originates from the fact that the sign
of $\lambda(W^{\pm})$ is flipped as compared to the case of $h^0$ while
$\lambda(\chi^{\pm}_i)$ remains positive. Fig.\,3 also shows the effects
of different $\eta$. (Figs.\,3 (a) and (c) vs. Figs.\,1 (a) and (c),
Figs.\,3 (b) and (d) vs. Figs.\,2 (a) and (c).) We verified the result of 
Ref.\cite{dicus} that the rate scales approximately as the logarithm of
$\eta$, for sufficiently small values of $\eta$.

Fig.\,4 serves as a comparison with the result of Ref.\cite{dicus} in which
the production of $A^0$ was calculated. The solid and dashed 
(dotted and dotdashed) curves are computed using the same input
parameters as in Figs.\,1 (b) ((c)) in that reference. We see that the 
cross section for $h^0$ is slightly larger than that for $A^0$ throughout
most of the range of $m_{A^0}$.

As a by-product, we show the production cross section for the SM Higgs boson 
$\phi^0$ \cite{ebo,cotti} in Fig.\,5.
The cross section for $50~{\rm GeV}<m_{\phi^0}<100~{\rm GeV}$ is comparable to
that of the MSSM $h^0$ in the same mass range and large 
$\tan\beta$ \cite{compar}.

In Figs.\,(1), (2) and (5), the short dashed lines represent the cross section
for the direct production of a $b\bar{b}$ pair with invariant mass $m_h$ in
$e\gamma$ collisions. To minimize this background, we impose an angular cut on 
the $b$ and $\bar{b}$ relative to the incident photon in the $e\gamma$ center 
of mass. For a cut of $|\cos\theta < .98\,|$ 
on both the $b$ and the $\bar{b}$, the background is reduced by a factor of
about 10 with the signal being essentially unchanged. For the MSSM light Higgs,
$h_0$, the background is reasonable, and at $\sqrt{s} = 500\,$ GeV and $m_{h^0}
 = 80\,$ GeV, the signal to square root of background ratio $S/\sqrt{B}$ is 4.5
for 50 fb$^{-1}$ of luminosity. 
The situation is less good for the MSSM heavy Higgs, $H^0$, where for
$\sqrt{s} = 500\,$ GeV and $m_{H^0} = 160\,$ GeV, $S/\sqrt{B}$ = 1.8. In the
case of the SM Higgs, $\phi_0$, the $b\bar{b}$ background is not significant
when $m_{\phi_0} > 75\,$GeV.

We have studied the production of neutral scalar Higgs bosons in the MSSM and
the SM at future electron colliders operating in the $e\gamma$ mode.
Although the results are potentially complicated by many free parameters
associated with supersymmetric particles, we find that the dominant 
contribution comes only from charginos, which is an automatic result in
the case of $A^0$ production due to its off-diagonal couplings to bosonic
particles$\cite{dicus}$. Furthermore, the chargino contribution is generally
significant and can be comparable to or even dominate over the ordinary
contributions in the case of $H^0$ production. This may help to identify a
particular type of new particle once there are experimental hints for
its existence in virtual loops. Finally, we comment that the 
cross sections computed here and in Ref.$\cite{dicus}$ are generally
one or two orders of magnitude larger than the related associated production
of photons and Higgs bosons in $e^+ e^-$ collisions \cite{djouadi}.
This implies that the $e\gamma$ channel may provide an important window in the 
search for Higgs bosons.
 
\acknowledgements

We are grateful to Y.-P. Kuang and C.-P. Yuan for helpful discussions and we are
indebted to Duane Dicus for providing us the data used in the background 
calculation. This work was supported in part by the National Natural Science 
Foundation of China under contract No. 19677205.

\appendix
\section{Reduced couplings $\lambda$ and $\lambda'$}

The following reduced couplings of the neutral scalar Higgs bosons
$h^0$ and $H^0$ in MSSM are obtained from the list of Feynman rules in
Ref.\cite{hunter}. The couplings for the SM Higgs boson $\phi^0$ are
recovered as a special case, $\lambda(W^{\pm})=\lambda(f)=1$ with the others
zero.

For the lighter Higgs boson $h^0$, we have
\begin{equation}
\begin{array}{rcl}
\lambda(W^{\pm}) & = & \sin(\beta-\alpha)\,,\\
\lambda(u) & = & \cos\alpha/\sin\beta\,,\\
\lambda(d) & = &-\sin\alpha/\cos\beta = \lambda(l)\,,\\
\lambda(H^{\pm}) & = &\sin(\beta-\alpha)
+\case{1}{2}\cos 2\beta \sin(\beta+\alpha)/\cos^2\theta_W\,,\\
\lambda(\chi^{\pm}_i) & = &\case{1}{2}[-(Q_{ii}+Q_{ii}^{\star})\sin\alpha
                                  +(S_{ii}+S_{ii}^{\star})\cos\alpha]\,,\\
\lambda'(\chi^{\pm}_i) & = &\case{1}{2}[(Q_{ii}-Q_{ii}^{\star})\sin\alpha
                                  -(S_{ii}-S_{ii}^{\star})\cos\alpha]\,,\\
\lambda(\tilde{u}_R) & = &Q_u \tan^2\theta_W \sin(\alpha+\beta)-
m^2_u\cos\alpha/m^2_W\sin\beta\,,\\
\lambda(\tilde{u}_L) & = &(\case{1}{2}-Q_u \sin^2\theta_W)
\sin(\alpha+\beta)/\cos^2\theta_W-m^2_u\cos\alpha/m^2_W\sin\beta\,,\\
\lambda(\tilde{d}_R) & = &Q_d\tan^2\theta_W\sin(\alpha+\beta)+
m^2_d \sin\alpha/m^2_W\cos\beta\,,\\
\lambda(\tilde{d}_L) & = &(-\case{1}{2}-Q_d\sin^2\theta_W)
\sin(\alpha+\beta)/\cos^2\theta_W+m^2_d\sin\alpha/m^2_W\cos\beta\,,
\end{array}
\end{equation}
while for the heavier Higgs, $H^0$,
\begin{equation}
\begin{array}{rcl}
\lambda(W^{\pm}) & = &\cos(\beta-\alpha)\,,\\
\lambda(u) & = &\sin\alpha/\sin\beta\,,\\
\lambda(d) & = &\cos\alpha/\cos\beta = \lambda(l)\,,\\
\lambda(H^{\pm}) & = &\cos(\beta-\alpha)
-\cos 2\beta\cos(\beta+\alpha)/2\cos^2\theta_W\,,\\
\lambda(\chi^{\pm}_i) & = &\case{1}{2}[(Q_{ii}+Q_{ii}^{\star})\cos\alpha
                                  +(S_{ii}+S_{ii}^{\star})\sin\alpha]\,,\\
\lambda'(\chi^{\pm}_i) & = &-\case{1}{2}[(Q_{ii}-Q_{ii}^{\star})\cos\alpha
                                  +(S_{ii}-S_{ii}^{\star})\sin\alpha]\,,\\
\lambda(\tilde{u}_R) & = &-Q_u \tan^2\theta_W \cos(\alpha+\beta)-
m^2_u\sin\alpha/m^2_W\sin\beta\,,\\
\lambda(\tilde{u}_L) & = &(-\case{1}{2}+Q_u \sin^2\theta_W)
\cos(\alpha+\beta)/\cos^2\theta_W-m^2_u\sin\alpha/m^2_W\sin\beta\,,\\
\lambda(\tilde{d}_R) & = &-Q_d\tan^2\theta_W\cos(\alpha+\beta)-
m^2_d \cos\alpha/m^2_W\cos\beta\,,\\
\lambda(\tilde{d}_L) & = &(\case{1}{2}+Q_d\sin^2\theta_W)
\cos(\alpha+\beta)/\cos^2\theta_W-
m^2_d\cos\alpha/m^2_W\cos\beta\,.
\end{array}
\end{equation}
$\lambda(\tilde{l}_R)$ and $\lambda(\tilde{l}_L)$ are obtained from 
$\lambda(\tilde{d}_R)$ and $\lambda(\tilde{d}_L)$ with $Q_d$ and $m_d$
replaced respectively by $Q_l$ and $m_l$. The $2\times 2$ matrices 
$Q_{ij}$ and $S_{ij}$ are defined in terms of unitary matrices $U$ and
$V$ that diagonalize the chargino mass matrix,
\begin{equation}
\displaystyle{
Q_{ij}=\frac{1}{\sqrt{2}}V_{i1}U_{j2},~~
S_{ij}=\frac{1}{\sqrt{2}}V_{i2}U_{j1}.
             }
\end{equation}
For details on $U$ and $V$, see Ref.\cite{haber}. The mixing between 
$\tilde{f}_R$ and $\tilde{f}_L$ is ignored except for the top squark.
The Feynman rules for the mass eigenstates $\tilde{t}_1$ and $\tilde{t}_2$
can be expressed in terms of the mixing angle $\phi_t$ and the Feynman
rules for $\tilde{t}_R$ and $\tilde{t}_L$. For all details on this, we
refer the reader to Refs.\cite{hunter,haber}.

%\newpage
\begin{figure}[h]
\hspace{0.9in}
\epsfysize=3.0in \epsffile{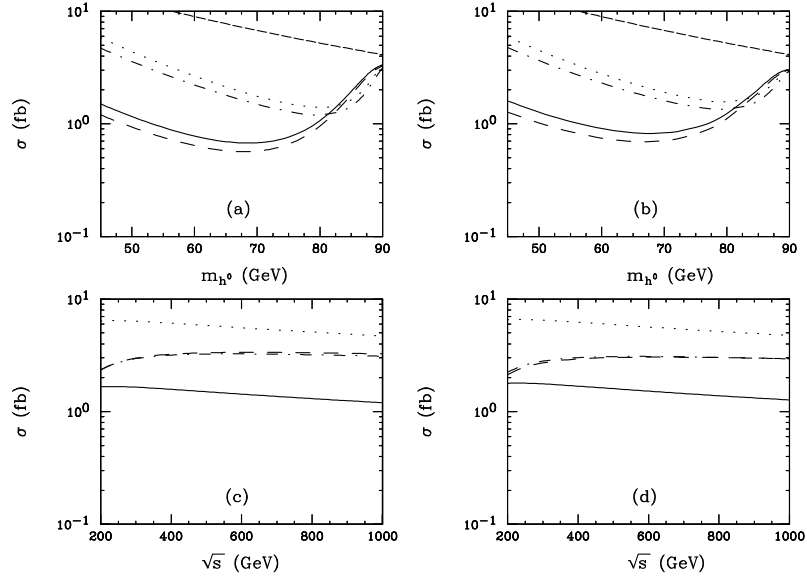}
\vspace{10pt}
\caption{The cross section for the production of the light Higgs boson $h^0$ 
is shown as a function of its mass $m_{h^0}$ and the center of mass energy
$\protect\sqrt{s}$. For panels (a) and (c), $M_{\protect\tilde{l},\protect
\tilde{\nu}}=
200$ GeV, $M_{\protect\tilde{q},\protect\tilde{u},\protect\tilde{d}}=500$ GeV, 
and $M=\mu=A_t=200$ GeV. In panel (a), the solid line corresponds to $\protect
\sqrt{s}=500$ GeV, $\tan\beta=15$,
the dotted to $\protect\sqrt{s}=500$ GeV, $\tan\beta=30$, the dashed to 
$\protect\sqrt{s}=
1000$ GeV, $\tan\beta=15$ and the dotdashed to $\protect\sqrt{s}=1000$ GeV, 
$\tan\beta=30$. In panel (c), the solid line corresponds to $m_{h^0}=45$ GeV, 
$\tan\beta=15$, the dotted to $m_{h^0}=45$ GeV, $\tan\beta=30$, the dashed to 
$m_{h^0}=90$ GeV, $\tan\beta=15$ and the dotdashed to $m_{h^0}=90$ GeV, 
$\tan\beta=30$. For panels (b) and (d), $M_{\protect\tilde{l},\protect
\tilde{\nu}}=500$ GeV, $M_{\protect\tilde{q},\protect\tilde{u},\protect
\tilde{d}}=500$ GeV, and $M=\mu=A_t=500$ GeV. The short dashed lines denote the
$b\protect\bar{b}$ background.}
\end{figure}
\begin{figure}[h]
\hspace{0.9in}
\epsfysize=3.0in \epsffile{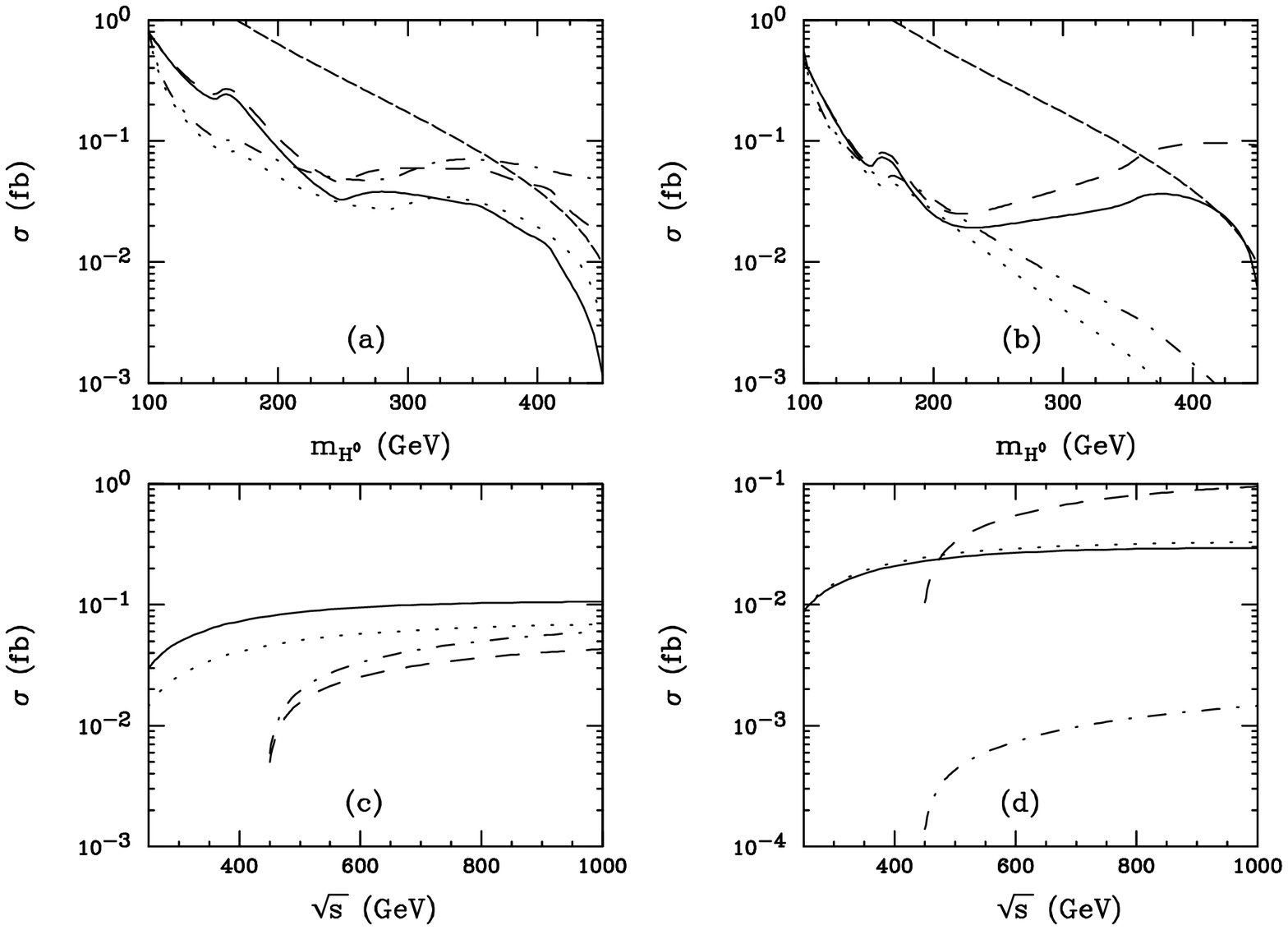}
\vspace{10pt}
\caption{The cross section for the production for the heavy Higgs boson $H^0$ 
is shown as a function of its mass $m_{H^0}$ and the center of mass energy 
$\protect\sqrt{s}$. For panels (a) and (c), $M_{\protect\tilde{l},\protect
\tilde{\nu}}=200$ GeV, $M_{\protect\tilde{q},\protect\tilde{u},\protect
\tilde{d}}=500$ GeV and $M=\mu=A_t=200$ GeV.
In panel (a), the solid line corresponds to $\protect\sqrt{s}=500$ GeV, 
$\tan\beta=2$,
the dotted to $\protect\sqrt{s}=500$ GeV, $\tan\beta=20$, the dashed to 
$\protect\sqrt{s}=
1000$ GeV, $\tan\beta=2$ and the dotdashed to $\protect\sqrt{s}=1000$ GeV, 
$\tan\beta=20$. In panel (c), the solid line corresponds to $m_{H^0}=200$ GeV, 
$\tan\beta=2$, the dotted to $m_{H^0}=200$ GeV, $\tan\beta=20$, the dashed to 
$m_{H^0}=400$ GeV, $\tan\beta=2$ and the dotdashed to $m_{H^0}=400$ GeV, 
$\tan\beta=20$. For panels (b) and (d), $M_{\protect\tilde{l},\protect
\tilde{\nu}}=500$ GeV, $M_{\protect\tilde{q},\protect\tilde{u},\protect
\tilde{d}}=500$ GeV, and $M=\mu=A_t=500$ GeV. The short dashed lines denote the
$b\protect\bar{b}$ background.}
\end{figure}
\begin{figure}[h]
\hspace{0.9in}
\epsfysize=3.0in \epsffile{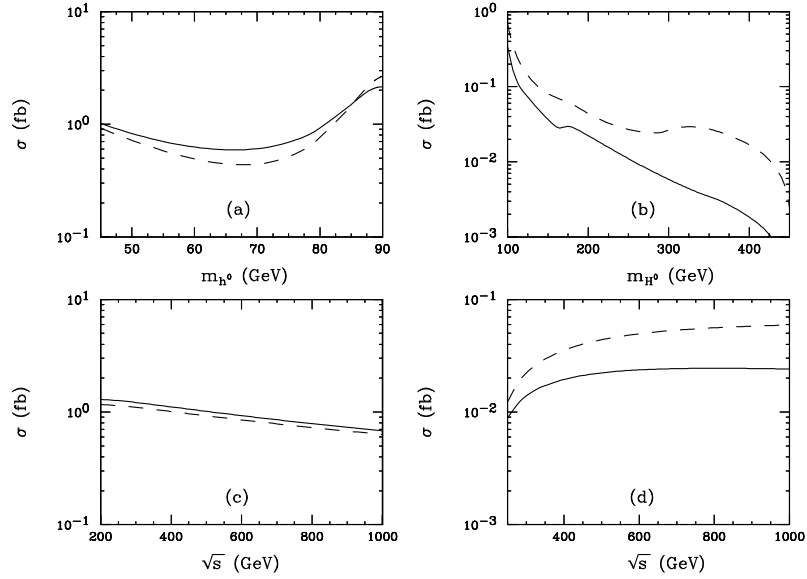}
\vspace{10pt}
\caption{The cross sections for the production of $h^0$ ((a) and (c)) and $H^0$ 
((c) and (d)) are shown as functions of their masses and $\protect\sqrt{s}$.
For all panels, the parameters are $M_{\tilde{l},\tilde{\nu}}=200$ GeV, 
$M_{\protect\tilde{q},\protect\tilde{u},\protect\tilde{d}}=500$ GeV and 
$M=\mu=A_t=200$ GeV, $\eta=10^{-4}$. In panel (a), $\protect\sqrt{s}=500$ GeV, 
$\tan\beta=15$; in (c) $m_{h^0}=45$ GeV, $\tan\beta=15$; in (b), 
$\protect\sqrt{s}=
500$ GeV, $\tan\beta=20$; and in (d), $m_{H^0}=200$ GeV, $\tan\beta=20$.
The solid curves are the contributions from $W^{\pm}$ and fermion loops,
while dashed curves include contributions from $\chi^{\pm}$.
The $H^{\pm}$ and scalar loops can be ignored.}
\end{figure}
\begin{figure}[h]
\hspace{.65in}
\epsfysize=2.8in \epsffile{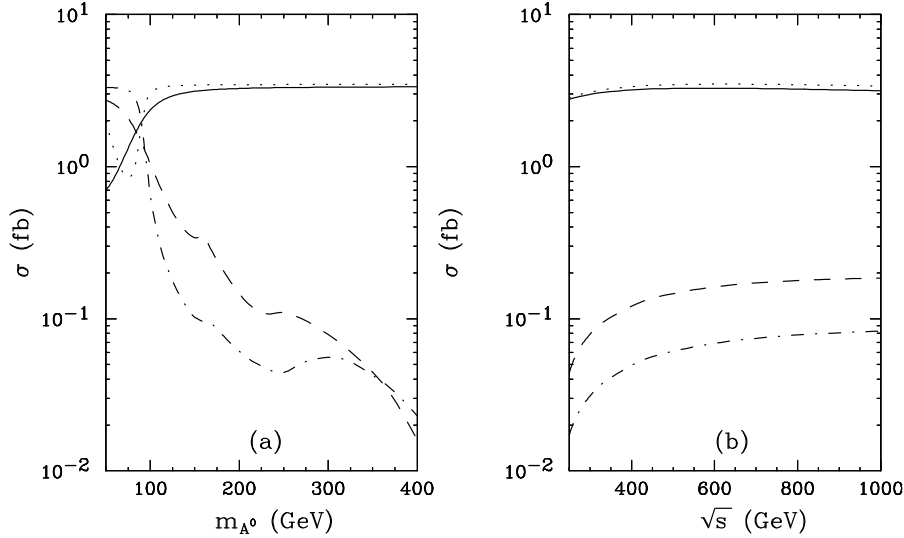}
\vspace{10pt}
\caption{The cross section for the production of $h^0$ 
(solid and dotted curves) and $H^0$ (dashed and dotdashed curves) are shown 
as functions of the mass $m_{A^0}$ of the pseudoscalar Higgs boson $A^0$
and $\protect\sqrt{s}$. For both panels, $M_{\protect\tilde{l},\protect
\tilde{\nu}}=
200$ GeV, $M_{\protect\tilde{q},\protect\tilde{u},\protect\tilde{d}}=500$ GeV 
and $A_t=200$ GeV. The solid and
dashed curves have $M=\mu=177$ GeV, $\tan\beta=5$, while the dotted and
dotdashed curves have $M=\mu=182$ GeV, $\tan\beta=20$. In panel (a), 
$\protect\sqrt{s}=
500$ GeV, and in (b), $m_{A^0}=200$ GeV.}
\end{figure}
\begin{figure}[h]
\hspace{.65in}
\epsfysize=2.8in \epsffile{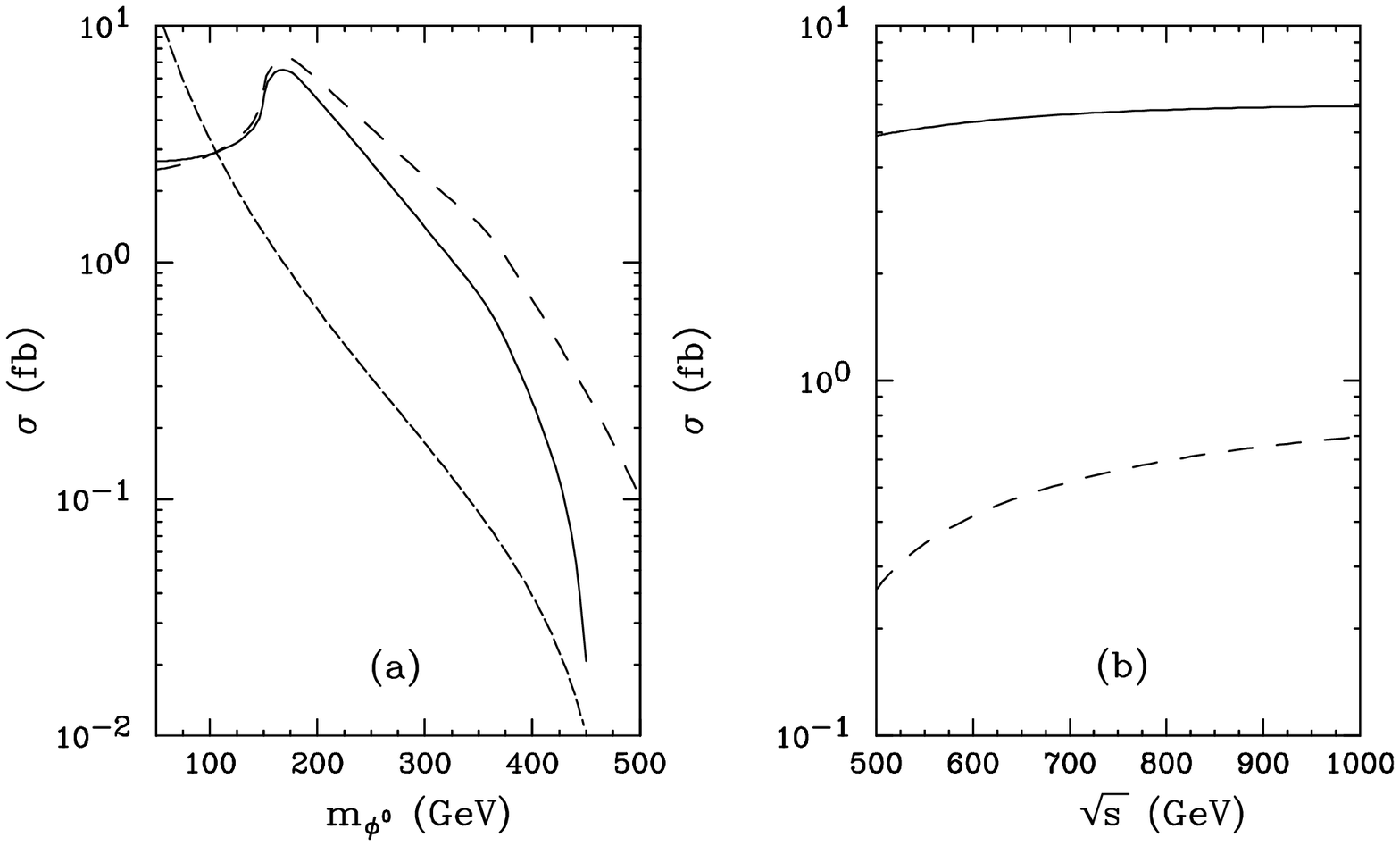}
\vspace{10pt}
\caption{The cross section for the production of the Standard Model Higgs 
boson $\phi_0$ is shown as a function of its mass $m_{\phi^0}$ and 
$\protect\sqrt{s}$.
In panel (a), the solid line is  $\protect\sqrt{s}=500$ GeV and the dashed is 
$\protect\sqrt{s}=1000$ GeV. In panel (b), the solid line is  $m_{\phi^0}=200$ 
GeV and the dashed is $m_{\phi^0}=400$ GeV. The short dashed line denotes the
$b\protect\bar{b}$ background.}
\end{figure}


\begin{thebibliography}{99}
\bibitem{compar} When our numerical work was being done, we became aware of  
Ref. \cite{cotti} in which the cross section for 
$e\gamma\to e\phi^0$ was computed in SM without using the $t$-channel
photon dominance. We obtain smaller results. The $W^{\pm}$
contribution in our Eq.\,(1) is consistent with the counterpart in
Eq.\,(12) of Ref.\cite{abbasabadi}, but differs from the one in 
Eq.\,(7) of Ref.\cite{cotti}.
\end{thebibliography}
\end{document}